\documentclass[]{pasj02}

\usepackage{natbib}
\usepackage{graphicx}
\usepackage{xcolor}
\usepackage{rotating}
\usepackage{booktabs}
\usepackage{adjustbox}
\usepackage{url}

\jyear{2026}
\Received{}
\Accepted{}

\begin{document}

\title{Remnant-wide mapping of heterogeneous presupernova burning histories in the O-rich ejecta of Cassiopeia A}

\author{
Yui \textsc{Kuboike},\altaffilmark{1}\altemailmark\orcid{0009-0000-8310-9463} \email{ce257009@meiji.ac.jp},
Toshiki \textsc{Sato},\altaffilmark{1}\altemailmark\orcid{0000-0001-9267-1693} \email{toshiki@meiji.ac.jp},
and
Kai \textsc{Matsunaga}\altaffilmark{2}\orcid{0009-0003-0653-2913} 
}

\altaffiltext{1}{Department of Physics, School of Science and Technology, Meiji University, Kanagawa, 214-8571, Japan}
\altaffiltext{2}{Department of Physics, Kyoto University, Kyoto 606-8502, Japan}

\KeyWords{stars: massive --- stars: evolution --- supernova remnants: individual (Cassiopeia A) --- nuclear reactions, nucleosynthesis, abundances --- X-rays: ISM}

\maketitle

\begin{abstract}
Late-stage shell interactions, including shell mergers, can mix material across adjacent burning shells and imprint chemically inhomogeneous structures in the hydrostatic O-rich layers before core collapse. Whether such inhomogeneities survive the explosion and remain observable in the resulting supernova remnant is not well understood. We investigate the spatial and compositional structure of the O-rich ejecta in Cassiopeia A using deep Chandra X-ray observations. Principal component analysis of narrow-band images reveals O-rich ejecta spanning a compositional sequence from Ne-rich through NeSi-rich to O/Si-rich material. Spatially resolved spectroscopy of 20 representative regions confirms systematic depletion of Ne and Mg toward the O/Si-rich component. Notably, the composition of the O/Si-rich ejecta over a broad elemental range from O to Ca is simultaneously well reproduced by O-layer material in 17--20 $M_\odot$ one-dimensional presupernova models that experienced strong shell-merger-induced burning and mixing. In contrast, the Ne-rich ejecta retain substantially higher Ne and Mg abundances and resemble less strongly processed O-shell material. These chemically distinct components are spatially segregated on remnant-wide scales, with Ne-rich ejecta preferentially concentrated toward the north and O/Si-rich ejecta more prominent in the southeast. Our results suggest that late-stage shell burning and mixing in the progenitor of Cassiopeia A were spatially heterogeneous. The remnant-wide distribution of ejecta with different burning histories further indicates that part of this asymmetric presupernova structure may have survived the explosion and remains imprinted in the present-day remnant.
\end{abstract}


\section{Introduction}
\label{sec:intro}

Late-stage shell burning in massive stars plays a crucial role in setting the initial conditions for core-collapse supernova explosions \citep{2002RvMP...74.1015W,2014ApJ...783...10S,2023ApJ...947...35B}. During the final stages of stellar evolution, vigorous convection develops in the nuclear-burning shells, generating large-scale fluctuations in velocity, density, and chemical composition. Recent multidimensional simulations have shown that these intrinsically three-dimensional progenitor structures can influence not only the subsequent explosion dynamics and the development of large-scale asymmetries, but also the transport and nucleosynthetic processing of the ejecta \citep{2017MNRAS.472..491M,2021ApJ...915...28B,2022MNRAS.510.4689V}.

In particular, shell mergers during the final stages of massive-star evolution have recently attracted considerable attention \citep{2007ApJ...667..448M,2011ApJ...733...78A,2020ApJ...890...94Y,2024MNRAS.533..687R}. In these events, an active burning shell interacts with neighboring shells and entrains chemically distinct material across the intervening stable layers. For example, an O-burning shell may incorporate Ne- and C-rich material from overlying layers. The ingestion of fresh fuel drives convective-reactive burning, enhances nuclear energy generation, and disrupts the compositional stratification between adjacent shells. Three-dimensional simulations show that this process generates vigorous large-scale flows and strongly asymmetric abundance distributions, rather than homogeneous mixing throughout the merged shell \citep{2020ApJ...890...94Y,2024MNRAS.533..687R}. The resulting velocity, density, and compositional perturbations may subsequently influence the core-collapse dynamics by producing anisotropic accretion onto the stalled shock and enhancing post-shock turbulence, thereby facilitating shock revival and contributing to explosion asymmetries \citep{2015ApJ...808L..21C,2017MNRAS.472..491M,2021ApJ...915...28B}.

Young supernova remnants provide a unique opportunity to test these predictions observationally, because both the spatial distribution and chemical composition of the stellar ejecta can be directly resolved. Cassiopeia A (Cas A), one of the youngest and brightest Galactic supernova remnants, is particularly well suited for this purpose. Its ejecta exhibit prominent large-scale asymmetries and chemically distinct structures, allowing the products of different burning layers to be studied across the remnant \citep{1979ApJ...233..154C,2010ApJ...725.2038D,2012ApJ...746..130H,2013ApJ...772..134M}. Recent X-ray spectroscopy has identified O-rich ejecta with unusually low Ne/Si and Mg/Si ratios, whose composition is consistent with O-shell material affected by late-stage shell merging in the progenitor \citep{2025ApJ...984..185S}. More recently, the detection of the odd-Z elements P, Cl, and K in O-rich ejecta has provided independent evidence that the progenitor experienced nonstandard late-stage burning and mixing, potentially including shell-merger-related nucleosynthesis \citep{2026NatAs..10..144X}. These observations suggest that Cas A preserves information about the presupernova burning structure, although the extent to which such structures survived or were rearranged by the explosion remains unclear.

Previous studies of possible shell-merger signatures in Cas A have generally been based on a small number of selected ejecta regions \citep{2025ApJ...984..185S}. It therefore remains unclear whether the previously identified Ne- and Mg-poor composition is representative of the O-rich ejecta as a whole or instead reflects localized burning and mixing within the progenitor. In particular, no remnant-wide study has yet established whether O-rich ejecta with different degrees of nuclear processing coexist within Cas A or whether they exhibit coherent large-scale spatial distributions. Furthermore, previous comparisons have largely emphasized a limited set of diagnostic elements, particularly Ne, Mg, and Si, although recent studies have begun to exploit a broader range of elements to investigate the burning history and ejecta structure of Cas A \citep{2025ApJ...995..155K,2025ApJ...990..103S}. A simultaneous comparison of O-rich ejecta over the full elemental range from O through S, Ar, and Ca nevertheless provides a more stringent test of whether the observed composition can be explained by a coherent presupernova shell-burning history.

In this study, we map the spatial and compositional diversity of the O-rich ejecta in Cas A using deep Chandra X-ray observations. We apply principal component analysis to narrow-band images of O, Ne, Mg, Si, and Fe to identify ejecta spanning a compositional sequence from Ne-rich to O/Si-rich material across the remnant. We then perform spatially resolved spectroscopy of representative regions and derive abundance ratios for elements from O through Ca. By comparing these measurements with one-dimensional presupernova models, we investigate whether the chemically distinct ejecta can be understood as O-shell material that experienced different degrees of late-stage burning and mixing. Our analysis is designed to test whether shell-merger-induced nucleosynthesis was spatially heterogeneous in the Cas A progenitor and whether this heterogeneous burning history remains imprinted on the large-scale structure of the remnant.

\section{Observation of Cas A} \label{observation}

To investigate the distribution of the O-rich layers across the entire Cas A remnant in greater detail, it is necessary to resolve the fine structures of the ejecta. Therefore, we used the deepest ACIS-S data targeting Cas A observed in 2004, with a total exposure of about 1 Ms. The level 1 event files were reprocessed using CIAO version 4.16 and CALDB 4.11.2 to apply the calibration products, including charge transfer inefficiency and time-dependent gain corrections, as well as filtering of bad grades and removal of pixel randomization. This paper employs a list of Chandra datasets, obtained by the Chandra X-ray Observatory, contained in DOI: \url{https://doi.org/10.25574/cdc.597}. 

\subsection{Principal Component Analysis}
\label{PCA}

\begin{figure*}[htbp]
\centering
\begin{tabular}{ccc}

\includegraphics[width=0.3\textwidth]{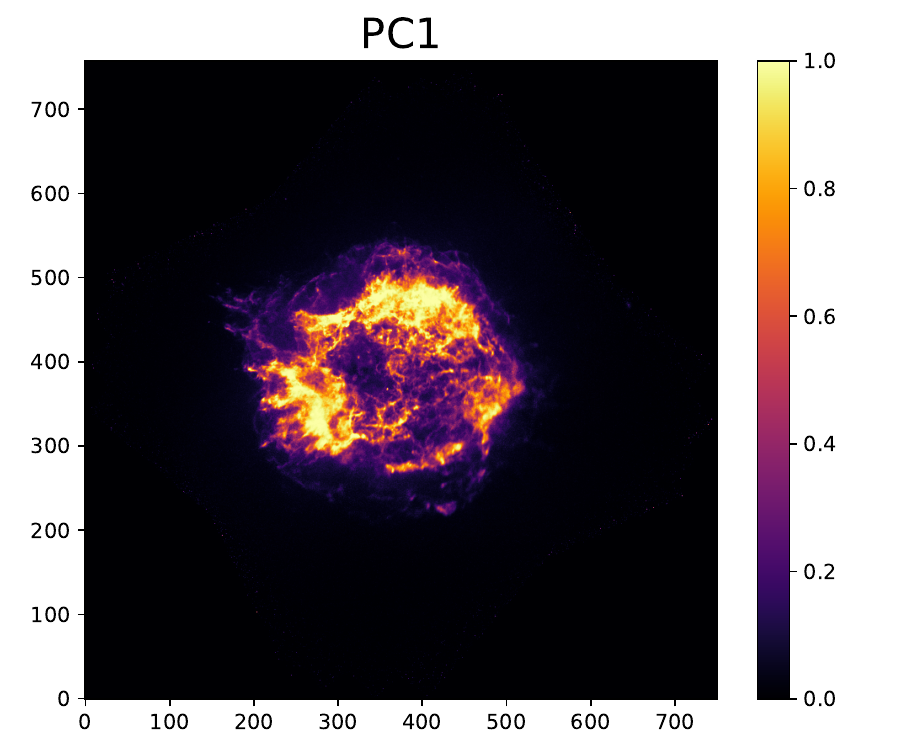} &
\includegraphics[width=0.3\textwidth]{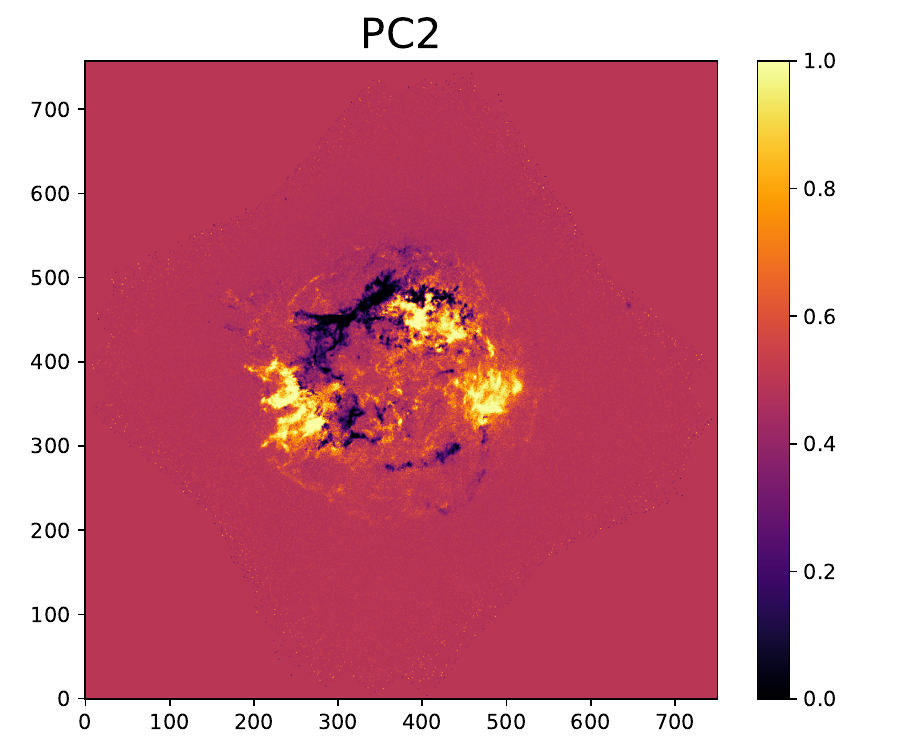} &
\includegraphics[width=0.3\textwidth]{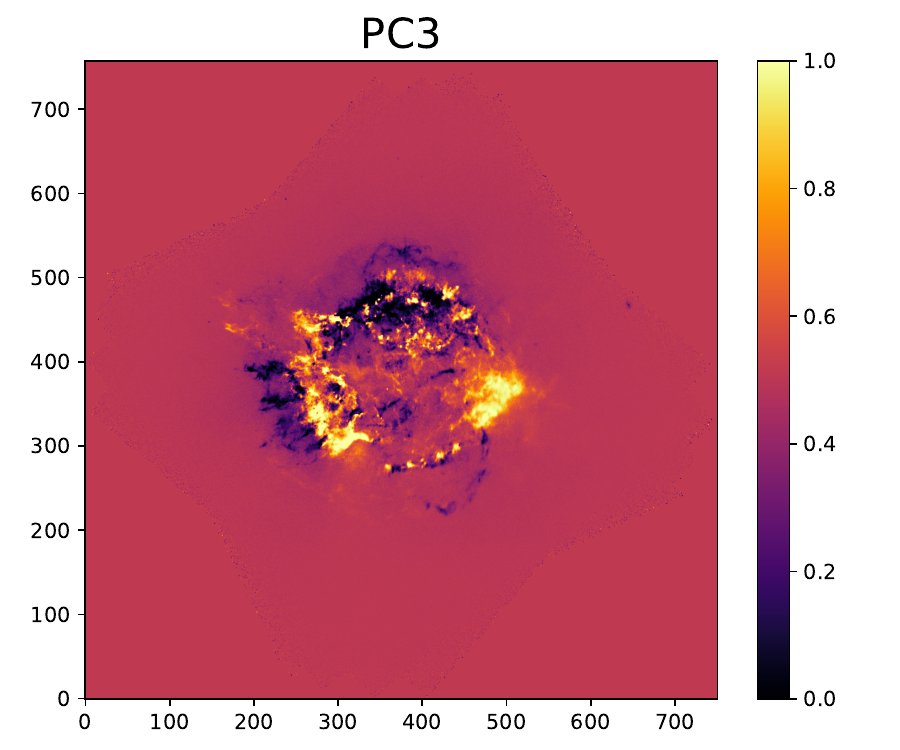} \\

\hspace{-10mm}\includegraphics[width=0.3\textwidth]{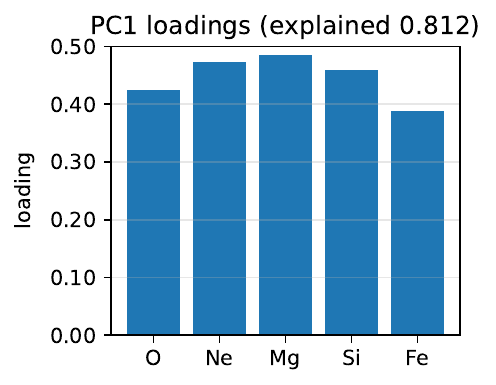} &
\hspace{-10mm}\includegraphics[width=0.3\textwidth]{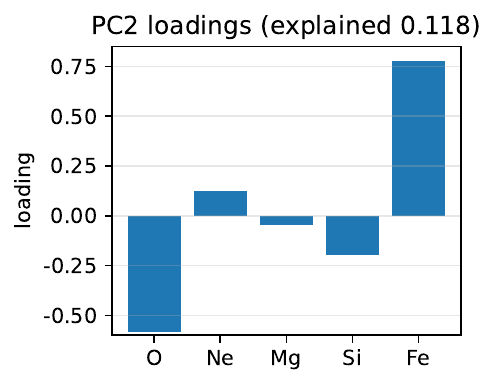} &
\hspace{-10mm}\includegraphics[width=0.3\textwidth]{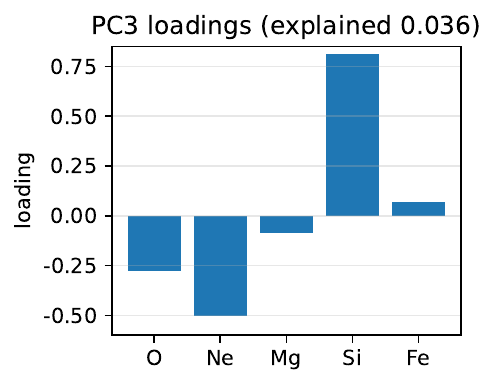} 

\\

\end{tabular}
\caption{Top: principal component maps (PC1--PC3). The color scale shows the transformed PC values, obtained by standardizing each principal component map, applying a hyperbolic tangent transformation, and rescaling the result to the range [0,1] for display. Bottom: corresponding PCA loadings which are the coefficients of the linear combination of the elemental-band fluxes that defines each component.}
\label{pca}
\end{figure*}

\begin{figure*}[htbp]
    \centering
    \includegraphics[width=0.9\linewidth]{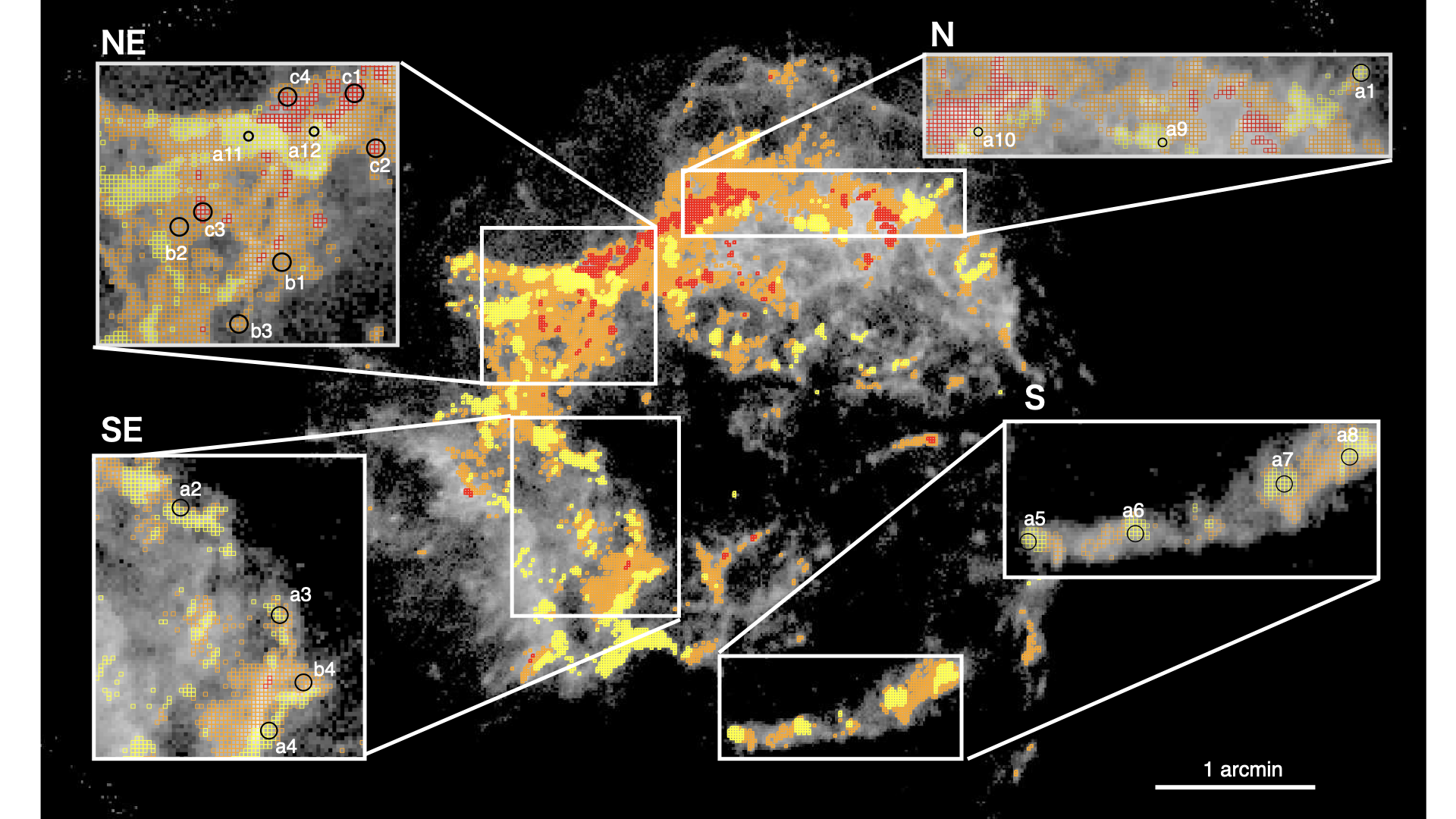}
    \caption{Spatial distribution of O-rich ejecta in Cas A identified through a principal component analysis. The ejecta are divided into three compositional classes: Si-rich (yellow), NeSi-rich (orange), and Ne-rich (red). Twenty regions were selected for spectral analysis. Labels “a,” “b,” and “c” correspond to the Si-rich, NeSi-rich, and Ne-rich components, respectively. The prefixes N, NE, SE, and S indicate the northern, northeastern, southeastern, and southern areas. The bright yellow knots (a9--a12) were extracted using circular regions with radii of $1''$, while all others use $2''$.}
    \label{20region}
\end{figure*}


To characterize compositional variations within the O-rich ejecta of Cas A, we applied principal component analysis (PCA) to narrow-band X-ray images tracing key elements. PCA enables the dominant patterns of correlated spectral variation to be identified and mapped across the remnant, and has previously been applied to spatially resolved X-ray studies of supernova remnants \citep{2005ApJ...634..376W}. In PCA, the input variables are first standardized, and their covariance matrix is decomposed into eigenvectors that define mutually orthogonal principal components. The data are then projected onto these components, which are ordered according to the fraction of the total variance that they explain. As input, we used flux maps of O (0.60--0.85 keV), Ne (0.98--1.10 keV), Mg (1.28--1.40 keV), Si (1.76--1.94 keV), and Fe (6.54--6.92 keV), which primarily trace K-shell emission from highly ionized species. The analysis was performed on a pixel-by-pixel basis using images of 751 $\times$ 758 pixels, with a pixel scale of 1 $\rm arcsec\, pixel^{-1}$.

The resulting principal-component maps and their corresponding loadings are shown in Figure \ref{pca}. The first three components account for 81.2\%, 11.8\%, and 3.6\% of the total variance, respectively. PC1 is dominated by positive contributions from all elemental bands and therefore primarily traces the overall X-ray surface brightness. In contrast, PC2 and PC3 capture secondary variations in the relative strengths of the elemental emission bands. PC2 mainly represents an anticorrelation between O and Fe emission, separating regions dominated by O-band emission from those with stronger Fe-K emission. PC3 primarily represents an anticorrelation between Ne and Si emission, tracing a compositional sequence from relatively Ne-bright to Si-bright O-layer material. Thus, the PCA decomposition provides an empirical means of identifying regions with systematically different X-ray line-emission properties.

We note that the narrow-band fluxes depend not only on elemental abundances but also on plasma temperature, ionization state, and foreground absorption, which cannot be independently disentangled by the PCA. We therefore use the PCA to identify regions with different relative line-emission properties, while their plasma conditions and elemental abundances are subsequently determined through spatially resolved spectral analysis in Section~\ref{spectral_analysis}.

To reduce the influence of extreme values while preserving small-scale spatial variations, each principal component map (PC1--PC3) was standardized using its mean and standard deviation. The maps were then transformed using a hyperbolic tangent function. This non-linear scaling compresses outliers more smoothly than a linear normalization and mitigates saturation effects. The transformed maps were then rescaled to the range [0,1] for visualization and further analysis.

\begin{figure*}[htbp]
    \centering
    \includegraphics[width=0.9\linewidth]{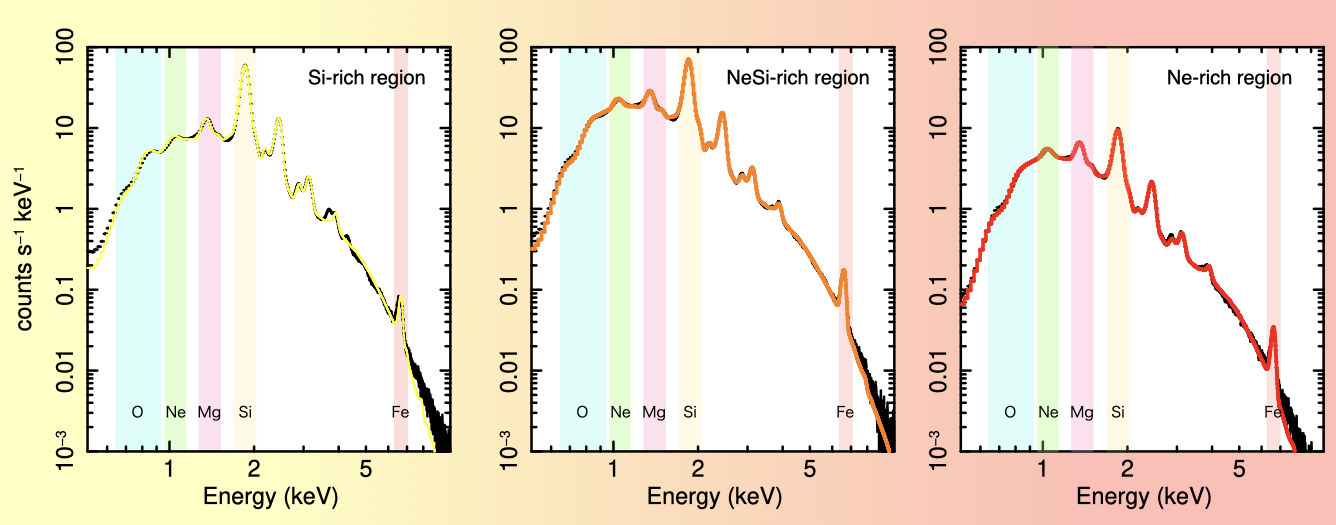}
    \caption{Best-fit spectral models for the integrated spectra extracted from the three PCA-classified regions shown in Figure~\ref{20region}. The yellow, orange, and red curves represent the spectra extracted from the yellow, orange, and red regions in Figure~\ref{20region}, respectively. Each spectrum was fitted with a thermal model (\texttt{phabs}$\times$(\texttt{vpshock}+\texttt{vpshock}+\texttt{vpshock})). The corresponding best-fit model and residuals are shown. }
    \label{3spectrum}
\end{figure*}

The ejecta regions were classified in the transformed PC2--PC3 parameter space. We first selected regions on the O-dominated side of the PC2 distribution by applying a threshold of $\mathrm{PC2}<0.3$, thereby excluding regions dominated by Fe-rich emission. Within this O-rich subset, three compositional classes were defined according to PC3: $\mathrm{PC3} < 0.01$, $0.01 < \mathrm{PC3} < 0.8$, and $\mathrm{PC3} > 0.8$. Based on the PC3 loadings and the corresponding elemental images shown in Figure~\ref{pca}, we refer to these classes as Ne-rich, NeSi-rich, and Si-rich ejecta, respectively. This interpretation is supported by the spectra and the derived abundance ratios of representative regions shown in Figure~\ref{3spectrum} and Table~\ref{3modelfit}. These divisions trace the systematic variation from Ne-rich to Si-enhanced compositions within the O-rich ejecta and form the basis of the regional classification.

Figure~\ref{20region} shows the spatial distributions of the three compositional classes, color-coded as red for Ne-rich, orange for NeSi-rich, and yellow for Si-rich ejecta. The Si-rich component is distributed relatively widely across the remnant, whereas the Ne-rich component is preferentially concentrated in the northern and northeastern regions. The O-rich ejecta therefore exhibit a pronounced large-scale compositional asymmetry. The physical origin of this asymmetry is discussed in Section \ref{discussion}.



\begin{figure*}[p]
    \centering
    \includegraphics[width=\linewidth]{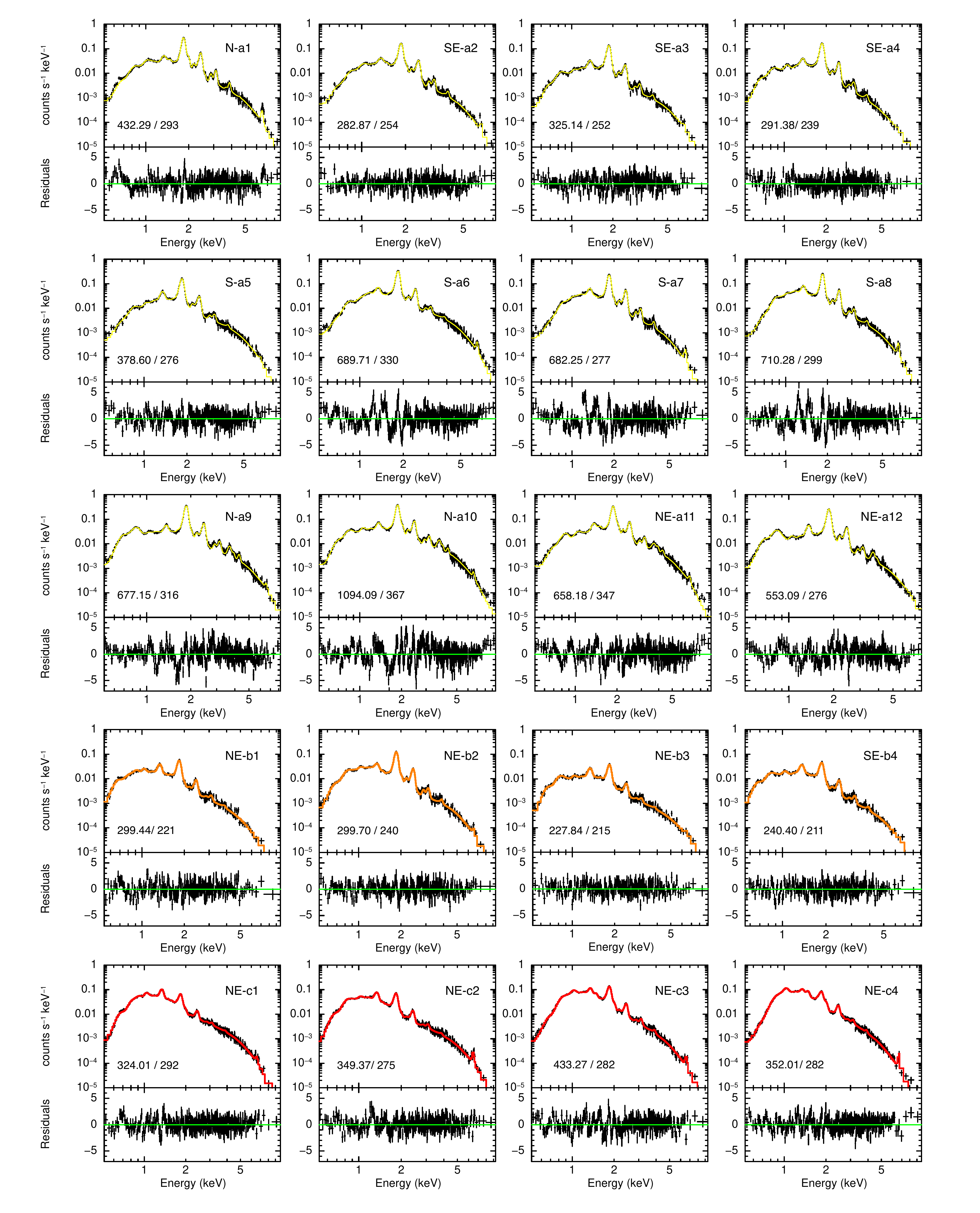}
    \caption{Spectra of the 20 regions shown in Figure~\ref{20region}. Each panel displays the observed spectrum, the best-fit plasma model, and the residuals. The color of the best-fit model indicates the compositional class: yellow for Si-rich, orange for NeSi-rich, and red for Ne-rich components. The region label is given in the upper right corner of each panel, and the $\chi^2/\mathrm{d.o.f.}$ value is shown in the lower left.
}
    \label{20spectrum}
\end{figure*}

\subsection{Spectroscopy of the PCA-classified Ejecta}
\label{spectral_analysis}

To quantify the elemental composition of the ejecta identified by the PCA classification, we performed spatially resolved spectroscopy of representative regions. Because the PCA is based on narrow-band flux images, it primarily captures relative spatial variations among the selected energy bands and does not by itself provide quantitative abundance measurements. We therefore derived abundance ratios from O through Fe for comparison with presupernova progenitor models in Section \ref{origin}. Spectral fitting was performed with XSPEC version 12.14.1 using AtomDB version 3.0.9 over the 0.5--10.0 keV energy range. Background spectra were subtracted from a nearby blank region.

From the three PCA-defined compositional classes, we selected 20 representative regions with minimal Fe contamination in order to isolate the O-rich ejecta as cleanly as possible. Because several bright knots in the Si-rich class are affected by pile-up, regions a9--a12 were extracted using circular apertures with radii of 1$^{\prime\prime}$, whereas all other regions were extracted using 2$^{\prime\prime}$ apertures. The spectra of regions a1--a8 and all regions in classes b and c were fitted with an absorbed two-component nonequilibrium-ionization plasma model, \texttt{phabs}$\times$(\texttt{vpshock}+\texttt{vpshock}). For regions a9--a12, two Gaussian components centered at approximately 3.7 and 4.3 keV were added to phenomenologically account for pile-up-related spectral features, giving the model \texttt{gaussian}+\texttt{gaussian}+\texttt{phabs}$\times$(\texttt{vpshock}+\texttt{vpshock}).

Figure~\ref{20spectrum} shows the spectra and best-fit models for all 20 regions. The PCA classification successfully separates the O-rich ejecta according to their dominant spectral characteristics. The Si-rich regions exhibit prominent Si K$\alpha$ emission in the 1.76--1.94 keV band, whereas the Ne-rich regions show stronger Ne and Mg emission at 0.98--1.10 keV and 1.28--1.40 keV, respectively. The NeSi-rich regions display intermediate spectral characteristics. The regions classified as Si-rich retain Si-dominated spectra even where the formal fit quality is relatively poor, including the southern regions a6--a8 and the pile-up-affected regions a9--a12.

\begin{table*}[htbp]
\centering
\tbl{Best-fit spectral parameters for three regions.\footnotemark[$*$]}{%
\begin{tabular}{lccc} \hline
Region & Si-rich & NeSi-rich & Ne-rich \\
\hline
$N_{\rm H}$ [$10^{22}\,\mathrm{cm}^{-2}$]
& $1.55\pm0.002$ & $1.35\pm0.001$ & $1.31\pm0.001$ \\
$kT_{1}$ [keV]
& $9.87\pm1.81$ & $6.85\pm0.16$ & $5.34\pm0.39$ \\
$kT_{2}$ [keV]
& $1.83\pm0.04$ & $3.50\pm0.03$ & $1.39\pm0.08$ \\
$kT_{3}$ [keV]
& $1.76\pm0.05$ & $1.07\pm0.003$ & $1.91\pm0.03$ \\
(O/H)/(O/H)$_\odot$
& $16.67\pm0.39$ & $11.39\pm0.17$ & $11.87\pm0.42$ \\
(Ne/H)/(Ne/H)$_\odot$
& $1.05\pm0.03$ & $2.69\pm0.04$ & $4.36\pm0.15$ \\
(Mg/H)/(Mg/H)$_\odot$
& $1.39\pm0.02$ & $1.50\pm0.01$ & $2.42\pm0.06$ \\
(Si/H)/(Si/H)$_\odot$
& $10.43\pm0.16$ & $5.87\pm0.06$ & $4.76\pm0.11$ \\
(S/H)/(S/H)$_\odot$
& $8.54\pm0.13$ & $5.55\pm0.05$ & $4.15\pm0.10$ \\
(Ar/H)/(Ar/H)$_\odot$
& $8.15\pm0.13$ & $6.19\pm0.06$ & $4.26\pm0.11$ \\
(Fe/H)/(Fe/H)$_\odot$
& $1.94\pm0.03$ & $1.00\pm0.01$ & $1.10\pm0.04$ \\
$\tau_{\rm u1}$ [$\mathrm{s\,cm^{-3}}$]
& $(5.03\pm0.66)\times10^{11}$ & $(9.02\pm0.10)\times10^{10}$ & $(1.11\pm0.03)\times10^{11}$ \\
$\tau_{\rm u2}$ [$\mathrm{s\,cm^{-3}}$]
& $(5.58\pm0.29)\times10^{10}$ & $(2.02\pm0.02)\times10^{10}$ & $(4.46\pm0.23)\times10^{10}$ \\
$\tau_{\rm u3}$ [$\mathrm{s\,cm^{-3}}$]
& $(1.08\pm0.12)\times10^{11}$ & $(3.12\pm0.04)\times10^{11}$ & $(1.25\pm0.08)\times10^{11}$ \\
Redshift $z$
& $(-1.13\pm0.001)\times10^{-2}$ & $(-7.98\pm0.001)\times10^{-3}$ & $(-7.77\pm0.006)\times10^{-3}$ \\
Norm 1 [$\frac{10^{-14}}{4\pi D^2}\int n_en_HdV\, \text{cm}^{-5}$]
& $(6.06\pm0.30)\times10^{-6}$ & $(3.48\pm0.09)\times10^{-5}$ & $(8.01\pm0.77)\times10^{-6}$ \\
Norm 2 [$\frac{10^{-14}}{4\pi D^2}\int n_en_HdV\, \text{cm}^{-5}$]
& $(3.23\pm0.61)\times10^{-2}$ & $(3.39\pm0.04)\times10^{-2}$ & $(9.73\pm0.67)\times10^{-3}$ \\
Norm 3 [$\frac{10^{-14}}{4\pi D^2}\int n_en_HdV\, \text{cm}^{-5}$]
& $(3.32\pm0.65)\times10^{-2}$ & $(1.18\pm0.01)\times10^{-1}$ & $(1.16\pm0.10)\times10^{-2}$ \\
$\chi^2$/d.o.f.
& $46908.54/631$ & $68933.96/631$ & $14419.39/566$ \\
\hline
\end{tabular}
}
\label{3modelfit}
\begin{tabnote}
\footnotemark[$*$] The solar values of \citet{1989GeCoA..53..197A} were used to estimate the mass ratios.
\end{tabnote}
\end{table*}

In the spectral fitting, the abundances of O, Ne, Mg, Si, S, Ar, and Fe were allowed to vary, while those of the remaining elements were fixed at their solar values. The Ca abundance was tied to Ar, and Ni was tied to Fe. The elemental abundances were tied between the two shock components, whereas their electron temperatures and upper ionization timescales were allowed to vary independently. The redshift was tied between the two components, and the lower ionization timescale of each \texttt{vpshock} component was fixed at zero. Parameter uncertainties are quoted at the $1\sigma$ confidence level when the reduced chi-square satisfies $\chi^2_\nu < 2$, accounting for variations in the other free parameters. For fits with $\chi^2_\nu > 2$, formal confidence intervals could not be robustly determined. We therefore report the approximate statistical errors provided by XSPEC as reference values only. These uncertainties should be regarded as approximate because they do not fully capture the systematic mismatch between the data and the adopted plasma model.

Although the PCA separates the O-rich ejecta along a clear sequence from Si-rich to Ne-rich compositions, highly ionized Fe K emission is not completely removed from all selected regions. For example, N-a1 and NE-c4 exhibit detectable Fe K emission. This residual contamination is expected because the PCA classification is based on narrow-band flux maps and captures relative spatial variations among the selected bands, rather than performing a full spectral decomposition of overlapping plasma components. 

As a complementary check on the regional analysis, we also extracted integrated spectra from each of the three PCA-defined compositional classes. The integrated spectra and their best-fit models are shown in Figure~\ref{3spectrum}, and the corresponding parameters are listed in Table \ref{3modelfit}. Because the integrated regions contain a broader range of plasma conditions and greater contamination from Fe-rich ejecta, we added a third \texttt{vpshock} component to the two-component model, adopting $\texttt{phabs}\times(\texttt{vpshock}+\texttt{vpshock}+\texttt{vpshock})$.
For this additional component, the Fe abundance was fixed at 1000 times the solar value to approximate an Fe-dominated ejecta plasma, while the abundances of the other elements were fixed at their solar values. This value is phenomenological and is not interpreted as a direct measurement of the Fe abundance. The lower ionization timescale of each component was fixed at zero; the remaining parameters are summarized in Table \ref{3modelfit}.

The abundance trends inferred from the integrated spectra are qualitatively consistent with those obtained from the 20 spatially resolved regions. However, the formal $\chi^2$ values are very large. This is primarily because each integrated spectrum combines emission from many regions with different temperatures, ionization ages, velocities, and elemental compositions, which cannot be fully represented by a small number of discrete plasma components. The high photon statistics also make the fits sensitive to relatively small systematic discrepancies between the data and the model. These discrepancies likely include uncertainties in the instrumental response, atomic data, and nonequilibrium-ionization plasma modeling. In particular, substantial residuals remain in the $\sim$1--2 keV band, where Ne lines overlap with the complex Fe-L emission and where uncertainties in the atomic modeling are especially important. We therefore use the integrated spectra primarily to confirm the overall abundance trends associated with the PCA-based classification, rather than as the principal source of precise quantitative abundance constraints.

\section{Discussion}\label{discussion}

Our PCA-based spatial classification and spatially resolved spectroscopy reveal a pronounced compositional asymmetry within the O-rich ejecta of Cas A. The ejecta exhibit a systematic compositional progression from Ne-rich material to Ne- and Mg-depleted, Si-enhanced material. In this section, we first compare the observed abundance patterns with the one-dimensional presupernova models of \citet{2018ApJ...860...93S} to investigate the burning history and internal chemical  structure of the progenitor's O-layer. Then we examine whether the compositional segregation can be interpreted as an imprint of spatially varying burning and mixing in the progenitor, while also considering the effects of asymmetric ejecta transport during and after the explosion.

\begin{table*}[htbp]
\centering
\tbl{Best-fit mass ratios for the 20 individual regions and the three integrated compositional classes.}{%
\vbox{%
\hbox{\resizebox{\textwidth}{!}{%
\begin{tabular}{lcccccccccccc}
\toprule
Region & N-a1 & SE-a2 & SE-a3 & SE-a4 & S-a5 & S-a6 & S-a7 & S-a8 & N-a9 & N-a10 & NE-a11 & NE-a12 \\
\midrule
O/Si
& $20.4^{+1.58}_{-13.0}$ & $19.6^{+3.05}_{-7.56}$ & $16.0^{+6.69}_{-8.09}$ & $18.1^{+6.90}_{-4.55}$ & $27.5\pm0.71$ & $27.8\pm13.4$ & $17.1\pm9.76$ & $23.5\pm9.73$ & $19.5\pm5.39$ & $30.6\pm10.7$ & $36.3\pm32.1$ & $24.5\pm20.0$ \\
Ne/Si
& $0.18^{+0.12}_{-0.11}$ & $0.21^{+0.03}_{-0.06}$ & $0.41^{+0.43}_{-0.23}$ & $0.57^{+0.35}_{-0.16}$ & $0.81\pm0.02$ & $0.17\pm0.15$ & $0.19\pm0.15$ & $0.56\pm0.26$ & $0.16\pm0.05$ & $0.62\pm0.20$ & $0.20\pm0.17$ & $0.42\pm0.34$ \\
Mg/Si
& $0.10^{+0.004}_{-0.06}$ & $0.14\pm0.04$ & $0.15^{+0.04}_{-0.06}$ & $0.14^{+0.07}_{-0.03}$ & $0.25\pm0.01$ & $0.15\pm0.06$ & $0.19\pm0.09$ & $0.23\pm0.08$ & $0.08\pm0.02$ & $0.11\pm0.03$ & $0.14\pm0.11$ & $0.15\pm0.12$ \\
S/Si
& $0.47\pm0.30$ & $0.39^{+0.06}_{-0.14}$ & $0.38^{+0.16}_{-0.15}$ & $0.42^{+0.12}_{-0.13}$ & $0.38^{+0.07}_{-0.01}$ & $0.36\pm0.13$ & $0.51\pm0.24$ & $0.50\pm0.17$ & $0.45\pm0.10$ & $0.41\pm0.12$ & $0.37\pm0.30$ & $0.41\pm0.31$ \\
Ar/Si
& $0.11^{+0.01}_{-0.07}$ & $0.09^{+0.01}_{-0.02}$ & $0.10^{+0.03}_{-0.04}$ & $0.12^{+0.03}_{-0.03}$ & $0.09\pm0.01$ & $0.09\pm0.03$ & $0.15\pm0.07$ & $0.13\pm0.05$ & $0.13\pm0.03$ & $0.12\pm0.04$ & $0.10\pm0.08$ & $0.10\pm0.07$ \\
Ca/Si
& $0.07^{+0.003}_{-0.05}$ & $0.06\pm0.01$ & $0.06^{+0.02}_{-0.01}$ & $0.08\pm0.02$ & $0.05\pm0.01$ & $0.05\pm0.01$ & $0.10\pm0.04$ & $0.08\pm0.02$ & $0.08\pm0.01$ & $0.08\pm0.02$ & $0.07\pm0.04$ & $0.06\pm0.03$ \\
Fe/Si
& $0.50^{+0.03}_{-0.32}$ & $0.53^{+0.16}_{-0.14}$ & $0.17^{+0.05}_{-0.07}$ & $0.02^{+0.03}_{-0.01}$ & $0.29^{+0.02}_{-0.01}$ & $0.92\pm0.25$ & $0.99\pm0.33$ & $0.89\pm0.21$ & $0.25\pm0.04$ & $0.25\pm0.05$ & $0.34\pm0.20$ & $0.01\pm0.01$ \\
\bottomrule
\end{tabular}%
}}%
\kern1.2em
\hbox{\resizebox{\textwidth}{!}{%
\begin{tabular}{lccccccccccc}
\toprule
Region & NE-b1 & NE-b2 & NE-b3 & SE-b4 & NE-c1 & NE-c2 & NE-c3 & NE-c4 & Si-rich & NeSi-rich & Ne-rich \\
\midrule
O/Si
& $25.4^{+5.0}_{-1.3}$ & $26.0^{+2.4}_{-10.4}$ & $41.6^{+7.6}_{-8.6}$ & $41.2^{+24.3}_{-16.8}$ 
& $75.5^{+5.2}_{-3.1}$ & $92.6^{+2.3}_{-6.2}$ & $45.4^{+14.4}_{-11.0}$ & $37.1^{+8.5}_{-7.6}$
&$21.83\pm0.61$ & $26.47\pm0.46$ & $34.03\pm1.45$ \\
Ne/Si
& $1.82^{+0.32}_{-0.09}$ & $0.89^{+0.20}_{-0.35}$ & $1.24^{+0.22}_{-0.40}$ & $1.51^{+0.46}_{-0.57}$ 
& $6.34^{+0.49}_{-0.72}$ & $3.10^{+0.09}_{-0.98}$ & $3.04^{+1.45}_{-0.91}$ & $7.93^{+1.10}_{-0.95}$
& $0.25\pm0.009$ & $1.14\pm0.02$ & $2.28\pm0.10$\\
Mg/Si
& $0.41^{+0.01}_{-0.02}$ & $0.19^{+0.03}_{-0.08}$ & $0.52^{+0.07}_{-0.06}$ & $0.51^{+0.22}_{-0.23}$ 
& $1.35^{+0.09}_{-0.04}$ & $0.88\pm0.20$ & $0.68^{+0.20}_{-0.18}$ & $1.30^{+0.16}_{-0.17}$
& $0.12\pm0.003$ & $0.24\pm0.003$ & $0.47\pm0.02$ \\
S/Si
& $0.43^{+0.04}_{-0.03}$ & $0.41^{+0.31}_{-0.16}$ & $0.38^{+0.04}_{-0.03}$ & $0.45^{+0.16}_{-0.19}$ 
& $0.40^{+0.03}_{-0.02}$ & $0.43^{+0.18}_{-0.02}$ & $0.39^{+0.14}_{-0.14}$ & $0.56^{+0.07}_{-0.05}$
& $0.43\pm0.009$ & $0.49\pm0.007$ & $0.45\pm0.01$ \\
Ar/Si
& $0.15\pm0.03$ & $0.09^{+0.01}_{-0.04}$ & $0.11^{+0.03}_{-0.03}$ & $0.17\pm0.06$ 
& $0.09^{+0.03}_{-0.02}$ & $0.14^{+0.02}_{-0.12}$ & $0.12^{+0.02}_{-0.10}$ & $0.11^{+0.02}_{-0.02}$
&$0.11\pm0.002$ & $0.15\pm0.002$ & $0.13\pm0.005$ \\
Ca/Si
& $0.10\pm0.02$ & $0.05^{+0.01}_{-0.02}$ & $0.07\pm0.02$ & $0.11\pm0.04$ 
& $0.06\pm0.02$ & $0.09\pm 0.01$ & $0.08^{+0.02}_{-0.01}$ & $0.07\pm0.01$
& $0.07\pm0.001$ & $0.10\pm0.001$ & $0.08\pm0.002$ \\
Fe/Si
& $0.00^{+0.00}_{-0.00}$ & $0.37^{+0.22}_{-0.16}$ & $0.41\pm0.07$ & $0.20^{+0.06}_{-0.09}$ 
& $1.17^{+0.14}_{-0.08}$ & $2.18^{+0.06}_{-0.05}$ & $1.42^{+0.46}_{-0.28}$ & $1.94^{+0.31}_{-0.23}$
& $0.49\pm0.007$ & $0.45\pm0.004$ & $0.61\pm0.01$ \\
\bottomrule
\end{tabular}%
}}%
}%
}
\label{20modelfit2}
\end{table*}

\subsection{Comparison with 1D Presupernova Models}
\label{comparison}

\begin{figure*}[htbp]
    \centering
    \includegraphics[width=0.3\linewidth]{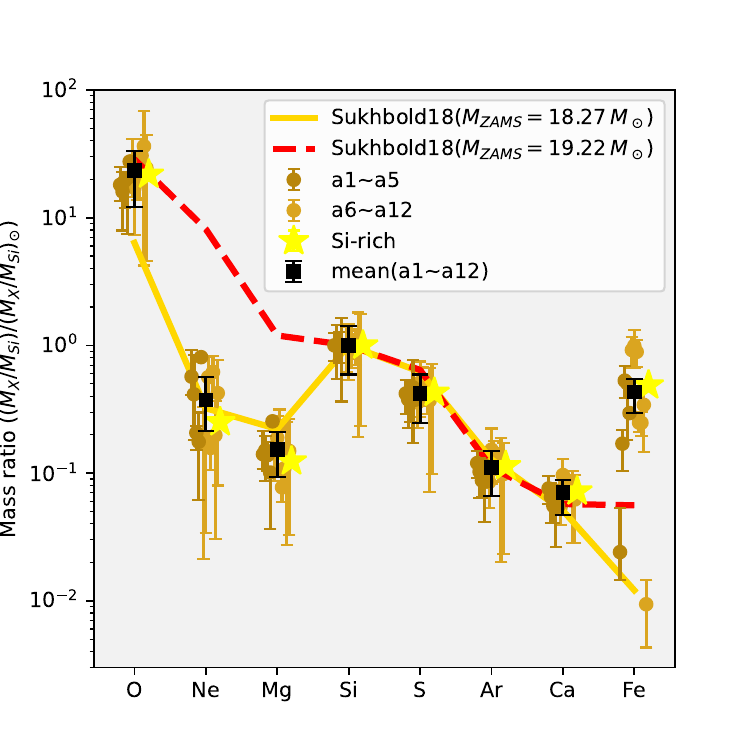}
    \includegraphics[width=0.3\linewidth]{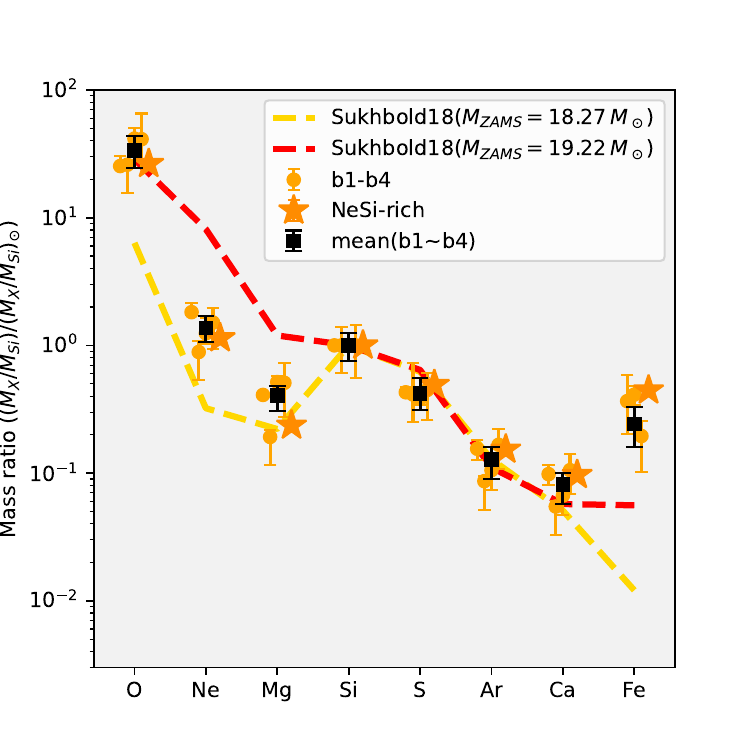}
    \includegraphics[width=0.3\linewidth]{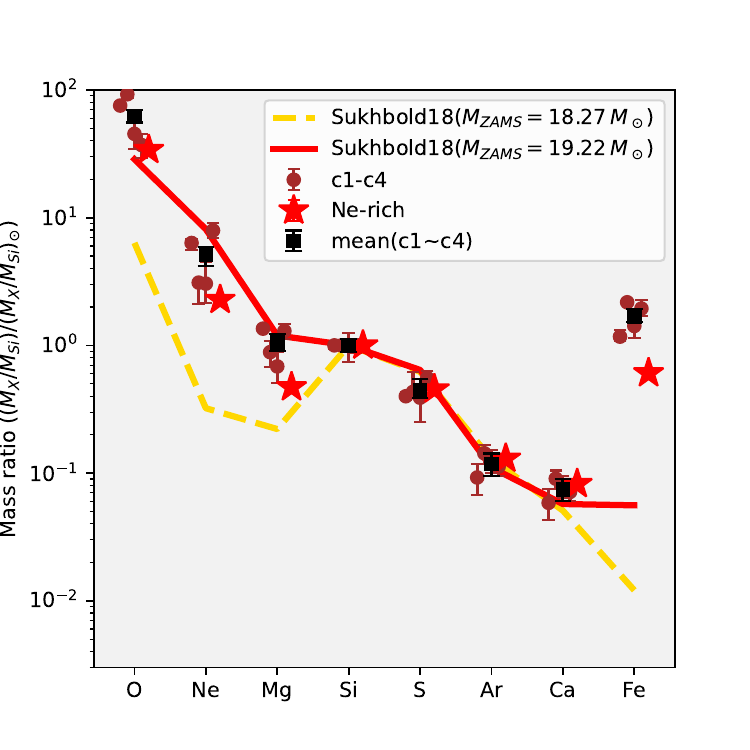}

    \vspace{0.2em}
    \includegraphics[width=0.3\linewidth]{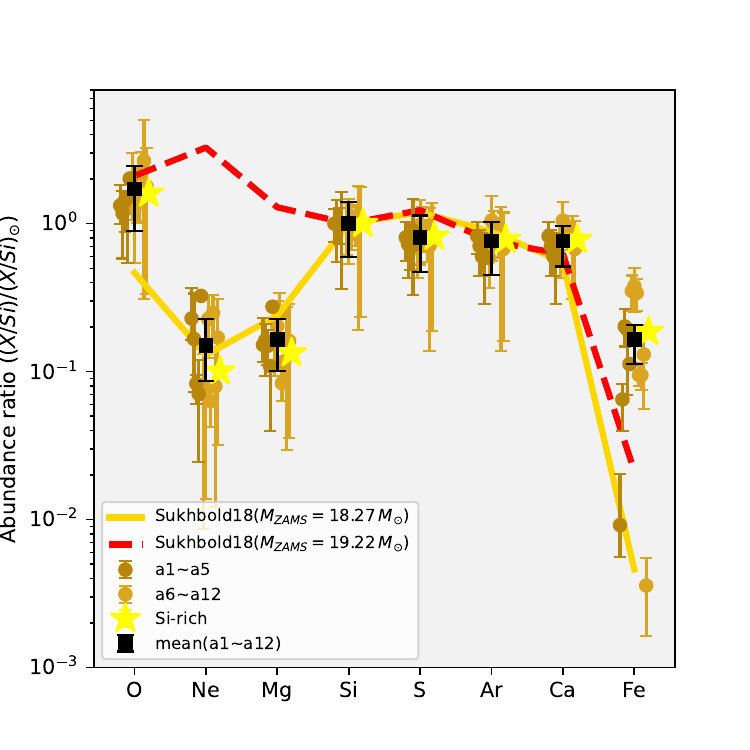}
    \includegraphics[width=0.3\linewidth]{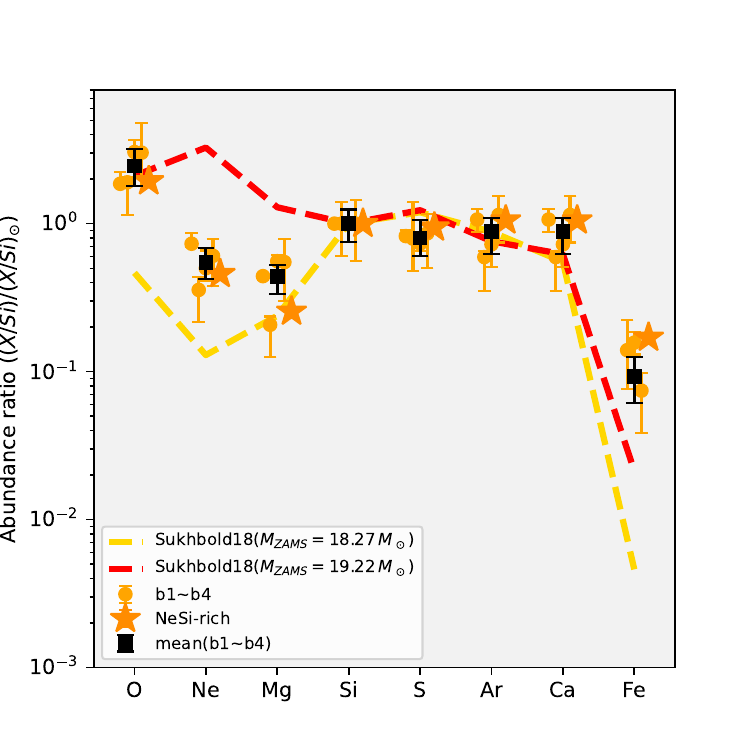}
    \includegraphics[width=0.3\linewidth]{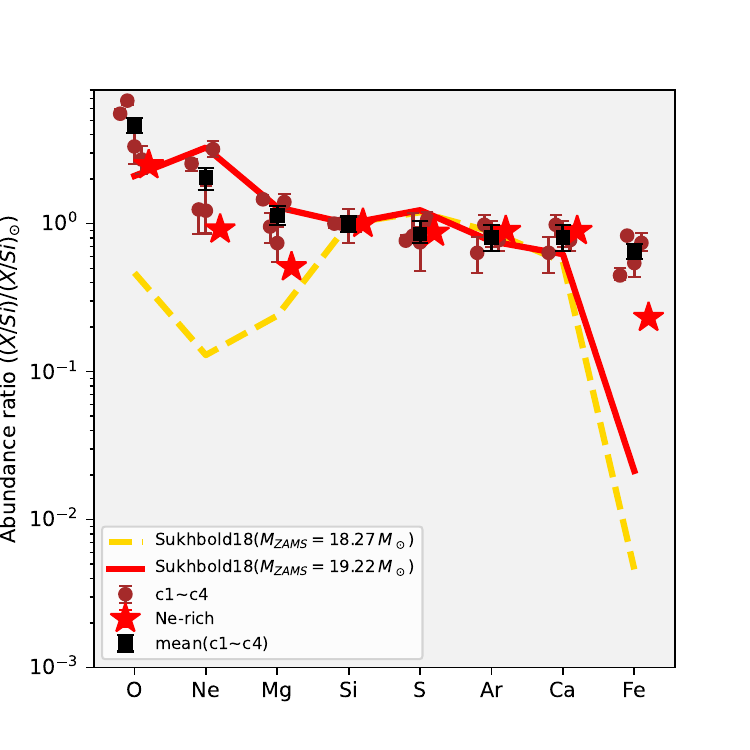}

    \caption{Mass ratios (upper panels) and elemental abundances (O, Ne, Mg, Si, S, Ar, Ca, and Fe; lower panels) normalized by Si of presupernova models \citep{2018ApJ...860...93S} and those of Cas A. 
    Left: Regions classified as Si-rich by the PCA analysis (a1--a12) are compared with the presupernova model that best reproduces the observed ratios, corresponding to a ZAMS mass of 18.27 $M_\odot$ (see text). Regions with $\chi^2\leq2$ (a1--a5) are shown in darker colors, while those with $\chi^2>2$ (a6--a12) are shown in lighter colors. Star symbols indicate the Si-rich values listed in Table \ref{3modelfit}. The black plots indicate the mean abundance of a1--a12. The red dashed curve shows the ZAMS model (19.22 $M_\odot$) that best matches the Ne-rich regions (c1--c4). Middle: The same presupernova models are compared with regions classified as NeSi-rich by the PCA analysis (b1--b4). Star symbols indicate the NeSi-rich values listed in Table \ref{3modelfit}. Right: Regions classified as Ne-rich by the PCA analysis (c1--c4) are compared with the presupernova model that best reproduces the observed ratios, corresponding to a ZAMS mass of 19.22 $M_\odot$. Star symbols indicate the Ne-rich values listed in Table \ref{3modelfit}.}
    \label{sukhboldplot}
\end{figure*}

Our spatially resolved spectroscopy provides quantitative abundance ratios for the compositionally distinct O-rich ejecta identified by the PCA analysis. Because O, Ne, and Mg in these ejecta are expected to preserve the presupernova O-layer composition more directly than elements synthesized predominantly during explosive Si burning, they provide useful diagnostics of the progenitor's late-stage shell evolution \citep{2025ApJ...984..185S}. In this section, we compare the abundance patterns of the Si-rich, NeSi-rich, and Ne-rich ejecta with the O-rich layers of the one-dimensional presupernova models of \citet{2018ApJ...860...93S}, selecting zones with an O mass fraction greater than 0.4 (i.e., O-rich layer). Rather than interpreting the best-matching models as independent determinations of the progenitor mass for each ejecta component, we use them to identify O-shell structures and burning histories that resemble the different compositions observed within Cas A. The corresponding ZAMS mass is reported because it provides the standard identifier for each model in the \citet{2018ApJ...860...93S} model grid, while our physical interpretation focuses on the associated O-shell structure and burning history. This comparison allows us to examine whether the Ne-rich and Si-enhanced O-rich ejecta can be understood as material that experienced different degrees of late-stage shell burning and mixing within a single progenitor.

The Si-rich, NeSi-rich, and Ne-rich classes contain 12 subregions (a1--a12), four subregions (b1--b4), and four subregions (c1--c4), respectively. For each subregion, we evaluate the agreement between the observations and the presupernova models by calculating the squared logarithmic residual,
\[
\left(\log A_{\rm obs}-\log A_{\rm model}\right)^2,
\]
for each element except Fe and summing the residuals over all included elements. We also tested a conventional $\chi^2$ minimization, but the result was strongly affected by a small number of subregions with exceptionally small statistical uncertainties. These uncertainties do not account for systematic uncertainties associated with the choice of plasma model and do not fully represent the uncertainties relevant to the model comparison. We thus avoided a statistical measure overly sensitive to them and adopted the unweighted logarithmic residual defined above. We then identify the best-matching model for each region as the one that minimizes the total residual and report its corresponding ZAMS mass.


Figure~\ref{sukhboldplot} compares the observed Si-normalized abundance and mass ratios with those predicted for the O-rich layers of the presupernova models. In addition to the results for the 20 individual regions, the ratios derived from the integrated spectra of the Si-rich, NeSi-rich, and Ne-rich classes are shown as star symbols in Figure~\ref{sukhboldplot}. These integrated values are broadly consistent with the distributions measured in the corresponding individual subregions. Table~\ref{20modelfit2} summarizes the ratios derived from both the 20 individual regions and the integrated spectra of the three compositional classes.

Among the models considered, the Si-rich regions (a1--a12) are best matched by the O-layer composition of the $M_{\rm ZAMS}=18.27\,M_{\odot}$ model. This model exhibits lower Ne/Si and Mg/Si ratios, representing an O-shell structure that has undergone stronger nuclear processing during late-stage shell burning and mixing, including shell-merger-induced burning \citep[e.g.,][]{10.1093/mnras/stx2470,2025ApJ...990..103S}. The Ne-rich regions (c1--c4) are best matched by the O-layer composition of the $M_{\rm ZAMS}=19.22\,M_{\odot}$ model, which retains a more typical O/Ne/Mg-rich composition without substantial depletion of Ne and Mg by late-stage shell burning. The NeSi-rich regions exhibit intermediate abundance patterns between these two classes. The systematic decrease in Ne/Si and Mg/Si from the Ne-rich toward the O/Si-rich regions can be interpreted as reflecting different degrees of C-shell ingestion into the O-burning shell and subsequent nuclear processing before core collapse. In contrast, S/Si, Ar/Si, and Ca/Si remain comparatively similar among these O-rich ejecta components.

A notable feature of the observed abundance pattern is the relatively high Fe abundance in many of the O-rich regions. In several regions, the Fe/Si ratios substantially exceed the values predicted for the presupernova O-rich layers, indicating an additional contribution from Fe-rich ejecta. Such material may have been mixed into the O-rich ejecta through hydrodynamic instabilities and large-scale ejecta overturn, or may be superposed along the line of sight \citep{2000ApJ...528L.109H,2012ApJ...746..130H}. The spatial overlap of some extraction regions with Fe K$\alpha$ emission in Figure~\ref{20region} is also consistent with this interpretation. Importantly, however, the presence of excess Fe does not appear to substantially alter the abundance pattern of the lighter elements. The systematic variation in Ne/Si and Mg/Si is preserved, while S/Si, Ar/Si, and Ca/Si remain comparatively similar among the O-rich ejecta components. This suggests that the excess Fe can be explained by a relatively small contribution from Fe-rich ejecta, without strongly modifying the underlying O-rich ejecta composition from O through Ca.

Some of the ejecta currently classified as O/Si-rich by the PCA may have originally had an O/Ne-rich composition and subsequently undergone explosive Ne and O burning during the supernova explosion, resulting in the depletion of O and Ne and a shift toward an O/Si-rich composition. Thus, the present-day PCA classification and spatial distribution do not necessarily provide a direct map of the presupernova composition. However, explosive Ne and O burning is expected to be confined to a relatively narrow region because of the rapid decline in post-shock temperature as the shock propagates outward \citep{1995ApJS..101..181W}. The observed variations in Ne/Si and Mg/Si are therefore likely to reflect predominantly compositional differences established during the presupernova evolution.

\begin{figure*}[htbp]
    \centering
    \includegraphics[width=\linewidth]{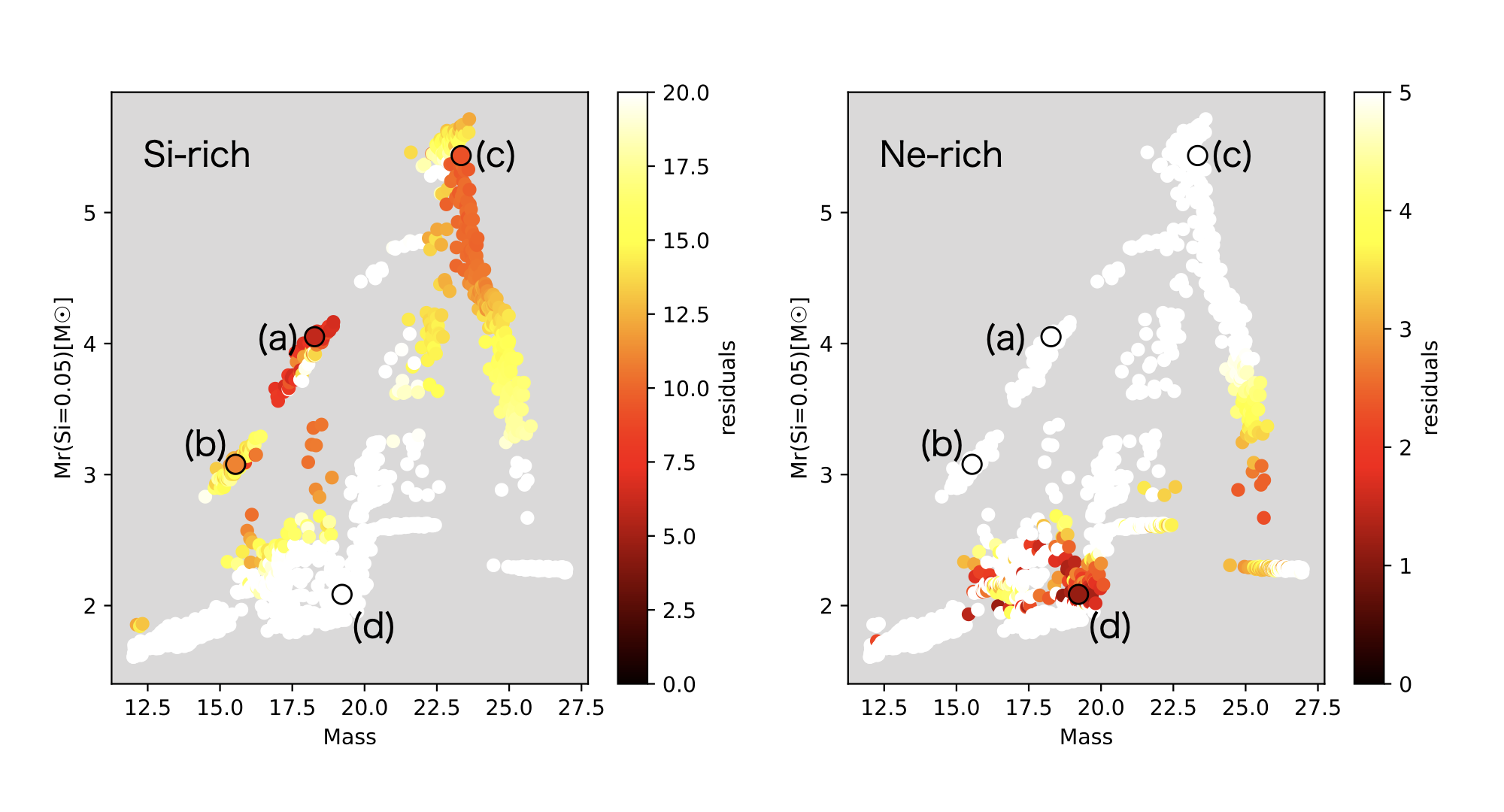}
    \caption{The extent of Si enrichment in the presupernova models as a function of ZAMS mass. The horizontal axis shows the progenitor ZAMS mass, and the vertical axis shows $M_r(\mathrm{Si}=0.05)$, defined as the outermost enclosed mass coordinate at which the Si mass fraction exceeds 0.05. Larger values of $M_r(\mathrm{Si}=0.05)$ indicate that Si-rich material extends farther outward into the O-rich shell and are characteristic of models with strong shell interaction or shell-merger-like mixing. Each point represents one of the one-dimensional presupernova models of \citet{2018ApJ...860...93S}. The point color indicates the total logarithmic residual between the observed abundance pattern and the model prediction, with darker colors corresponding to smaller residuals and therefore better agreement. Left: residuals calculated for the Si-rich ejecta. The best-matching models are concentrated primarily at $M_{\rm ZAMS}\sim17$--$20\,M_{\odot}$ and relatively large $M_r(\mathrm{Si}=0.05)$, indicating extended Si enrichment of the O-rich shell. Right: residuals calculated for the Ne-rich ejecta. These regions preferentially match models with smaller $M_r(\mathrm{Si}=0.05)$, in which the O-rich shell retains more Ne and is less enriched in Si.
    }
    \label{si-out}
\end{figure*}

\begin{figure*}[htbp]
    \centering
    \includegraphics[width=\linewidth]{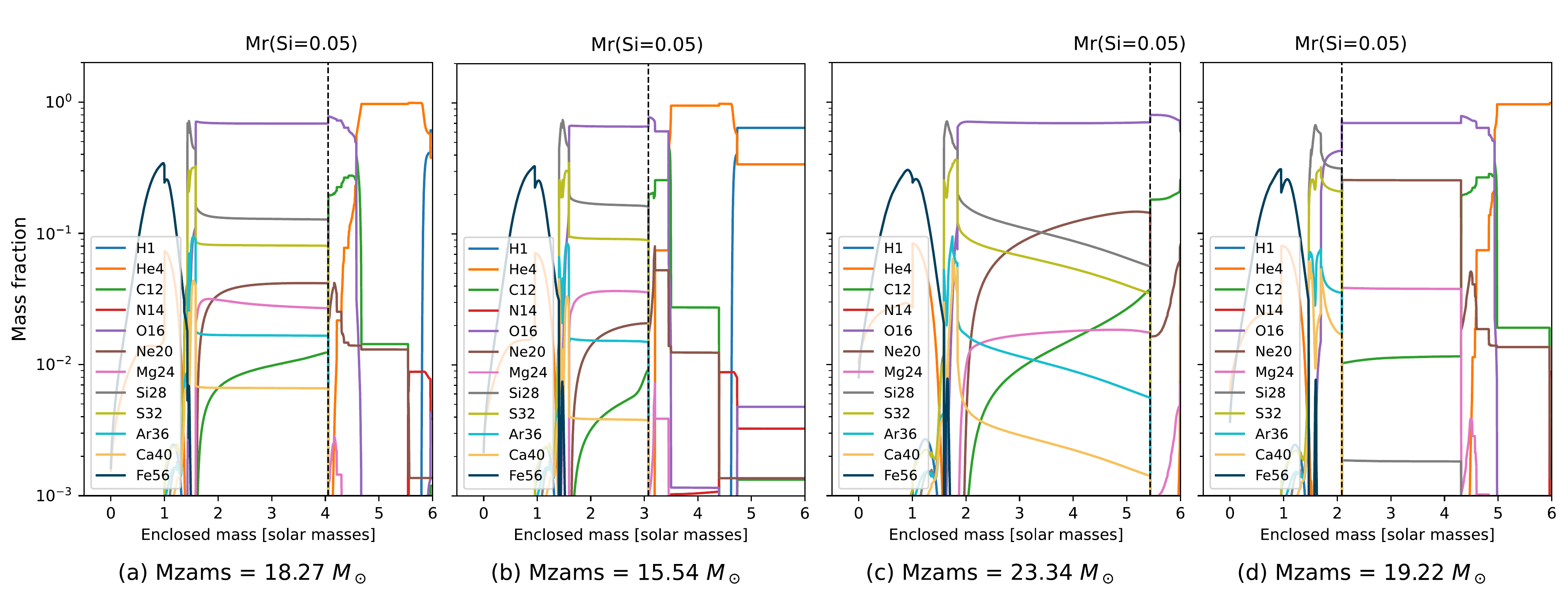}
    \caption{Mass-fraction profiles of four representative one-dimensional presupernova models selected from Figure~\ref{si-out}. Panels (a)--(d) correspond to ZAMS masses of $18.27$, $15.54$, $23.34$, and $19.22\,M_{\odot}$, respectively. The $18.27\,M_{\odot}$ model in panel (a), which most closely matches the Si-rich ejecta, exhibits strong Si enrichment and marked Ne and Mg depletion in the O-rich shell. The lower-mass model in panel (b) has a thinner O-rich shell and predicts more severe Ne depletion, whereas the higher-mass model in panel (c) has a more extended O-rich shell with Ne increasing and Si decreasing outward. The $19.22\,M_{\odot}$ model in panel (d), which most closely matches the Ne-rich ejecta, retains substantial Ne and Mg and contains relatively little Si in the O-rich shell. }
    \label{massfrac_profile4}
\end{figure*}

To further investigate the progenitor structure, we examine the model residuals as a function of the parameter $M_r(\mathrm{Si}=0.05)$ for the one-dimensional presupernova models of \citet{2018ApJ...860...93S}
\citep[see also][]{2025ApJ...984..185S,2025ApJ...995..155K}. Here, $M_r(\mathrm{Si}=0.05)$ is defined as the outermost enclosed mass coordinate at which the Si mass fraction exceeds 0.05.
This parameter provides a useful diagnostic of the radial extent of Si enrichment within the progenitor: a larger value indicates that Si-rich material extends farther outward into the O-rich shell.

In models that retain a relatively undisturbed onion-like structure, the Si-rich material remains confined to smaller enclosed mass coordinates.
By contrast, strong interactions or mergers between the O-burning shell and the overlying Ne- and C-rich shells can transport Si-rich material outward while processing Ne- and Mg-bearing material, thereby increasing $M_r(\mathrm{Si}=0.05)$.
Within the model set considered here, values of $M_r(\mathrm{Si}=0.05)\gtrsim3\,M_\odot$ are therefore characteristic of O-C shell-merger-like O-shell structures 
\citep{2025ApJ...984..185S,2025ApJ...995..155K}.
This value should be regarded as an empirical diagnostic for these particular presupernova models rather than as a universal threshold for O-C shell merging.

For each observed region, we compare the logarithmic residuals defined above with the corresponding $M_r(\mathrm{Si}=0.05)$ values of the models.
The results are shown in Figure~\ref{si-out}, where the color scale from red to black represents progressively smaller residuals and hence closer agreement between the observed Si-normalized abundance pattern and the model prediction.
This comparison allows us to determine whether the models favored by the observed composition are preferentially associated with extended, O-C shell-merger-like Si distributions in the O-rich shell.

These results indicate that the Si-rich regions are preferentially reproduced by a group of presupernova models with ZAMS masses of approximately $17$--$20\,M_{\odot}$, including the $18.27\,M_{\odot}$ model labeled (a). 
These models are characterized by relatively large values of $M_r(\mathrm{Si}=0.05)$, indicating that Si-rich material extends outward into the O-rich shell. 
Lower-mass models in the $15$--$17\,M_{\odot}$ range, represented by model (b), and higher-mass models above $20\,M_{\odot}$, represented by model (c), also show moderate agreement with the Si-rich observations. 
In contrast, the Ne-rich regions are preferentially matched by models with smaller $M_r(\mathrm{Si}=0.05)$ values, such as model (d), in which the O-rich shell retains more Ne and contains less Si.

To examine the physical origin of these trends, we compare the mass-fraction profiles of representative models from each group. 
Although individual models show numerical variations, the selected profiles illustrate the characteristic differences among the O-shell structures favored by the observations. 
Figure~\ref{massfrac_profile4} shows the mass-fraction profiles of models (a)--(d).

The $18.27\,M_{\odot}$ model, which provides the closest match to the Si-rich regions, exhibits strong Si enrichment throughout a substantial portion of the O-rich shell, accompanied by marked depletion of Ne and Mg. 
This composition resembles O-shell material that experienced strong late-stage shell-merger-induced burning and mixing. 
The lower-mass $15.54\,M_{\odot}$ model shown in panel (b) has a thinner O-rich shell and predicts more severe Ne depletion than is generally observed in the Si-rich ejecta. 
This comparison demonstrates that the residual Ne abundance provides an important constraint on the degree of nuclear processing in the O-rich shell.

The higher-mass $23.34\,M_{\odot}$ model in panel (c) contains a more extended O-rich shell. 
Within this shell, the Ne abundance increases outward while the Si abundance decreases, indicating that the outer part of the shell experienced less complete nuclear processing. 
Although this model reproduces some aspects of the Si-rich observations, its strong radial compositional gradient differs from the more uniformly Si-enhanced structure of model (a).

In contrast, the $19.22\,M_{\odot}$ model shown in panel (d), which best matches the Ne-rich regions, retains substantial Ne and Mg in the O-rich shell and contains relatively little Si. 
The contrast between models (a) and (d) therefore represents two distinct O-shell compositions: strongly processed, Si-enhanced material and more weakly processed, Ne-retaining material. 
Because both types of ejecta are observed within Cas A, the different best-matching models should not be interpreted as independent determinations of the progenitor mass. 
Rather, they suggest that compositional structures resembling those produced by different one-dimensional presupernova models coexisted within a single progenitor. 
This result supports a picture in which shell-merger-induced burning and mixing proceeded spatially nonuniformly in the Cas A progenitor, leaving different portions of the O-rich shell with distinct nuclear-burning histories.

\subsection{Origin of the O-rich Ejecta Asymmetry}
\label{origin}

\begin{figure*}[htbp]
    \centering
    \includegraphics[width=0.9\linewidth]{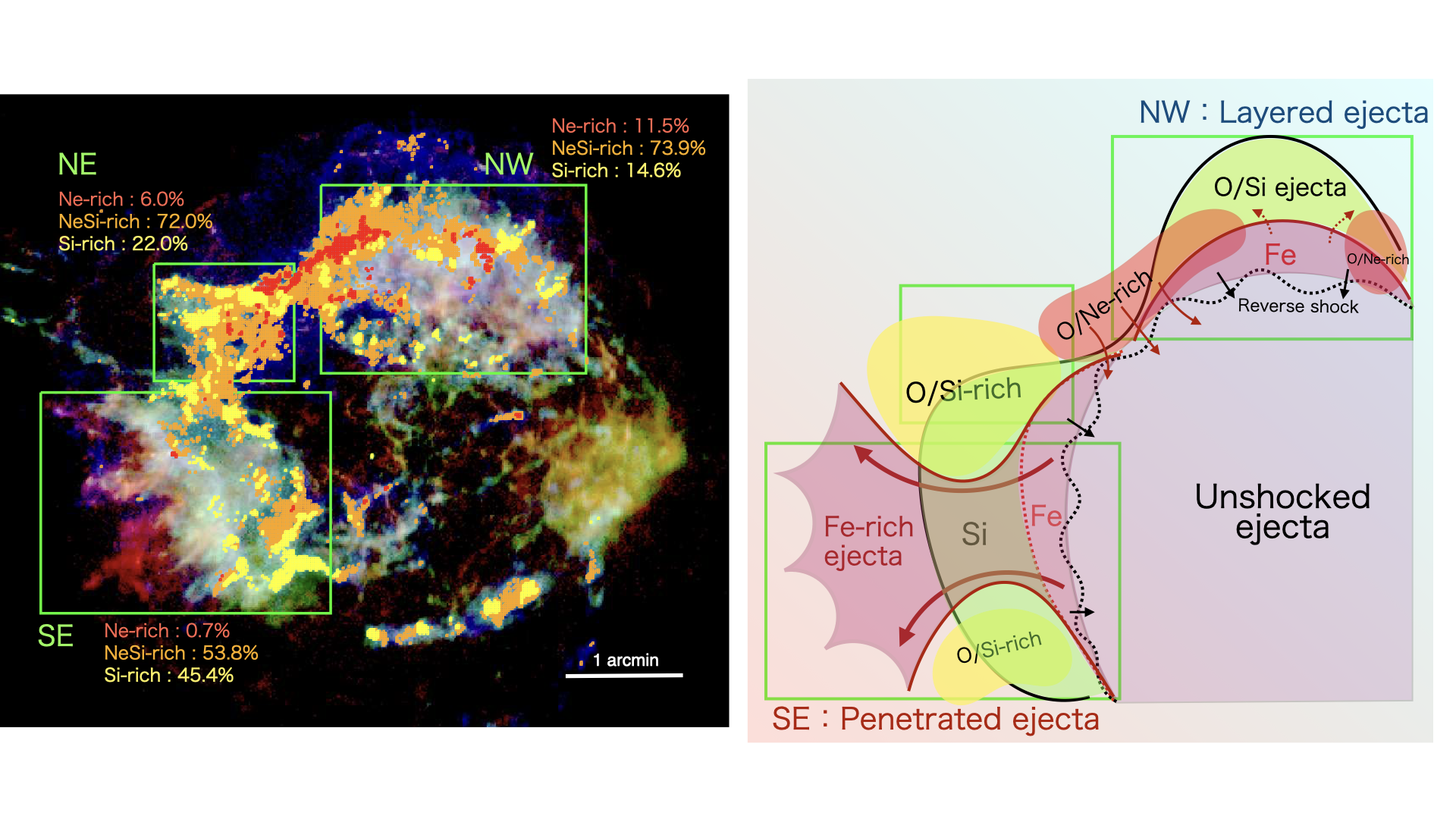}
    \caption{Spatial distribution of elements in Cas A. Left: RGB image of the elemental emission, overlaid with the PCA-based classification of the O-rich ejecta described in Section \ref{observation}. Red, green, and blue represent emission in the 6.54--6.92 keV (Fe He$\alpha$), 1.76--1.94 keV (Si He$\alpha$), and 0.60--0.85 keV (O He$\alpha$) bands, respectively. The rectangular regions labeled NW, NE, and SE enclose three prominent ejecta structures. For each region, we calculate the fractions of O-rich pixels classified as Si-rich, NeSi-rich, and Ne-rich. Right: Schematic illustration of a possible interpretation of the large-scale ejecta structure in Cas A. The colors follow those in the left panel: Fe-rich emission is shown in red, Si-rich emission in green, O/Si-rich ejecta in yellow, and O/Ne-rich ejecta in red-orange. The diagram illustrates spatially varying shell-merger-induced burning in the progenitor and subsequent asymmetric transport of the ejecta during the explosion. The reverse shock propagates inward in the ejecta frame and heats the different plasma components.}
    \label{NWNESE}
\end{figure*}

The large-scale asymmetry of the ejecta in Cas A has often been interpreted primarily in terms of asymmetric explosion dynamics, particularly for the Fe-rich and other innermost ejecta components \citep[e.g.,][]{2000ApJ...528L.109H,2013ApJ...772..134M,2014Natur.506..339G,2015Sci...347..526M,2021Natur.592..537S,2023ApJ...954..112S,2026ApJ..1001...46S}. The compositional asymmetry newly identified within the O-rich ejecta, however, may also retain information about the structure of the progenitor before core collapse. In particular, the spatial segregation between Ne-rich and Si-enhanced O-rich material may reflect spatial variations in the degree of late-stage burning and mixing within the progenitor's O shell. We therefore examine whether the observed distribution can be understood as a partial imprint of progenitor-stage convection and shell-merger-induced burning, while also considering redistribution of the ejecta during the explosion and subsequent remnant evolution.

To quantify this large-scale compositional variation, we compare the PCA classification of the O-rich ejecta with the elemental flux maps shown in Figure~\ref{NWNESE} (left). We define three rectangular regions encompassing the prominent northwestern (NW), northeastern (NE), and southeastern (SE) ejecta structures. Within each region, we calculate the fractions of O-rich pixels classified as Si-rich, NeSi-rich, and Ne-rich. The SE region has the largest fraction of Si-rich pixels, 45.4\%, compared with 29.0\% in the NE and 14.6\% in the NW. Conversely, the Ne-rich component is most prominent in the NW, where it accounts for 11.5\% of the classified O-rich pixels. These differences demonstrate a pronounced directional variation in the composition of the O-rich ejecta.

Based on the PCA classification and the elemental distribution maps, we propose a possible scenario for the origin of the large-scale compositional asymmetry, as illustrated schematically in Figure~\ref{NWNESE} (right). In this picture, large-scale convective flows developed within the progenitor's O-rich shell before core collapse, while the efficiency of fuel entrainment and nuclear burning varied strongly with direction. Three-dimensional shell-merger simulations indeed predict vigorous large-scale flows and strongly asymmetric, sometimes dipolar, abundance distributions within the merged shell \citep{2020ApJ...890...94Y,2024MNRAS.533..687R}. 

In the NE--NW regions, material carried by outward-moving convective flows may have retained an O/Ne-rich composition because the ingestion of Ne- and C-rich fuel into the active burning region was relatively inefficient. 
Some of this less processed material could then have been redirected by compensating inward flows and accumulated near the base of the Si/O-rich layer, allowing substantial amounts of Ne and Mg to survive until core collapse. 
In contrast, the SE may have experienced more vigorous fuel entrainment and convective-reactive burning. Three-dimensional simulations show that Ne-rich material entrained into the O-burning shell can be consumed, while Si-rich material produced by O burning is mixed outward by convection \citep{2020ApJ...890...94Y}. In this region, Ne- and Mg-rich material incorporated into the O-burning shell may have been processed more extensively, producing a Ne- and Mg-depleted, Si-enhanced O-rich composition.

During the subsequent explosion, a buoyant, high-entropy Fe-rich plume may have penetrated the overlying layers and entrained or displaced the strongly processed O/Si-rich material outward, as suggested by the Fe-rich plume structures observed in Cas A \citep{2021Natur.592..537S}. Large-scale perturbations inherited from presupernova shell burning may also have contributed to the development of the explosion asymmetry \citep{2017MNRAS.472..491M}. Lateral motions around the rising plume could then have displaced the O/Si-rich material to both sides of the Fe- and Si-rich structure, qualitatively producing the distribution observed today. Although this scenario is not unique, it provides a physically motivated connection between spatially heterogeneous shell-merger-induced burning in the progenitor and asymmetric ejecta transport during the explosion.

Another notable feature is the apparent deficit of O-rich ejecta in the western part of Cas A. The O-rich ejecta components identified and spectroscopically characterized in this study are distributed mainly from the SE through the NE to the NW, whereas no comparable structures are clearly detected in the western region. This apparent deficit may partly be an observational selection effect, because the western side of Cas A is viewed through a relatively large absorbing column, which strongly suppresses the soft X-ray emission used to trace O emission \citep[e.g.,][]{2012ApJ...746..130H}. Spatial variations in reverse-shock heating may also affect its X-ray visibility, although the interaction of the western ejecta with relatively dense circumstellar or interstellar material suggests that delayed shock heating alone is unlikely to account for the observed deficit \citep[e.g.,][]{2018ApJ...853...46S,2022ApJ...929...57V,2022A&A...666A...2O}.

If the western deficit is intrinsic, it would imply that the asymmetry extends beyond spatial variations in the burning history and also affects the global distribution of ejecta originating from the progenitor's O-rich shell. Such a structure could arise from strongly anisotropic shell burning in the progenitor, asymmetric ejecta transport during the explosion, or a coupling between the two. Distinguishing this intrinsic asymmetry from the observational effects discussed above is essential for determining how directly the present-day ejecta morphology retains the structure of the progenitor's O-rich shell.

Future X-ray microcalorimeters combining high spectral resolution with sufficient effective area below $\sim1$ keV would provide an important test. They can separate weak O emission lines from the continuum and neighboring Fe-L features and distinguish plasma components with different velocities and ionization states. Such observations could reveal faint O-rich plasma toward the highly obscured western region and determine whether its apparent deficit is caused primarily by absorption and plasma-selection effects or instead represents a genuine large-scale asymmetry in the distribution of the O-rich ejecta.

\section{Summary and Conclusions}

We investigated the spatial and compositional structure of the O-rich ejecta in Cas A using deep Chandra ACIS-S observations. 
Principal component analysis of narrow-band images tracing O, Ne, Mg, Si, and Fe isolates the dominant compositional variations within the remnant and identifies three classes of O-rich ejecta: Ne-rich, NeSi-rich, and Si-rich components. 
Spatially resolved spectroscopy of 20 representative regions confirms that these classes form a systematic compositional sequence, with Ne/Si and Mg/Si decreasing toward the Si-rich ejecta, while S/Si, Ar/Si, and Ca/Si remain comparatively uniform.

Comparison with the one-dimensional presupernova models of \citet{2018ApJ...860...93S} shows that the Si-rich ejecta are preferentially reproduced by models with ZAMS masses of approximately $17$--$20\,M_{\odot}$ and relatively large values of $M_r(\mathrm{Si}=0.05)$. 
These models contain O-rich shells in which Si-rich material extends to large enclosed mass coordinates and Ne and Mg are strongly depleted, as expected from substantial shell interaction and shell-merger-induced burning. 
In contrast, the Ne-rich ejecta resemble models with smaller $M_r(\mathrm{Si}=0.05)$, in which the O-rich shell retains substantially more Ne and Mg and contains less Si.

The different best-matching model masses should not be interpreted as independent estimates of the Cas A progenitor mass. Instead, they represent distinct one-dimensional O-shell structures that resemble different ejecta components observed within the same remnant. 
Their coexistence therefore suggests that strongly processed, Si-enhanced material and more weakly processed, Ne-retaining material were both present within a single progenitor.

The three compositional classes also exhibit markedly different large-scale spatial distributions.
Ne-rich ejecta are preferentially concentrated in the northern and northeastern regions, whereas Si-enhanced O-rich ejecta are more prominent in the southeast. When combined with the distinct O-shell structures inferred from the model comparison, this spatial segregation suggests that the progenitor's O-rich shell was not processed uniformly before core collapse.
One possible interpretation is that the efficiency of fuel entrainment and convective-reactive burning varied strongly with direction during a late-stage shell merger. 
The resulting presupernova compositional asymmetry may subsequently have been coupled to asymmetric ejecta transport during the explosion, including the development of Fe-rich plumes. 
The apparent paucity of O-rich ejecta in the western region may indicate an additional large-scale asymmetry, although strong foreground absorption and spatial variations in reverse-shock heating currently prevent a definitive interpretation.
Taken together, these results provide observational evidence that shell-merger-induced burning and mixing in the Cas A progenitor were spatially heterogeneous and that part of this heterogeneous burning history may have survived the explosion and remains imprinted on the present-day remnant.

Future soft-X-ray observations combining high spectral resolution with sufficient effective area below $\sim 1$ keV will be important for testing this interpretation. 
Such observations could identify faint O-rich plasma in the highly absorbed western region, resolve velocity and ionization components within the different compositional classes, and determine whether the observed asymmetry primarily reflects the presupernova burning structure, asymmetric ejecta transport, or subsequent remnant evolution.

\begin{ack}
This work was partly supported by Japan Society for the Promotion of Science Grants-in-Aid for Scientific Research (KAKENHI) Grant Number P23K13128 (TS) and 24KJ1485(KM).
During the preparation of this manuscript, the authors used ChatGPT (OpenAI; GPT-5.6) to assist with English-language editing, rewriting, and code refinement. All AI-assisted outputs were reviewed and, where necessary, modified by the authors, who take full responsibility for the content of the manuscript.
\end{ack}


\bibliography{sample631}{}
\bibliographystyle{aasjournalv7}

\end{document}